\documentclass{article}

\PassOptionsToPackage{numbers, compress}{natbib}

\usepackage[final]{neurips_ai4d3_2023}

\usepackage[utf8]{inputenc} 
\usepackage[T1]{fontenc}    
\usepackage{hyperref}       
\usepackage{url}            
\usepackage{booktabs}       
\usepackage{amsfonts}       
\usepackage{nicefrac}       
\usepackage{microtype}      
\usepackage{xcolor}         
\usepackage{graphicx}       

\usepackage{subcaption}
\usepackage{enumitem}
\usepackage{amssymb}
\usepackage{algorithm} 
\usepackage{algpseudocode} 

\usepackage{amsmath,amsfonts,bm}

\def\eqref#1{equation~\ref{#1}}

\def\1{\bm{1}}

\DeclareMathAlphabet{\mathsfit}{\encodingdefault}{\sfdefault}{m}{sl}
\SetMathAlphabet{\mathsfit}{bold}{\encodingdefault}{\sfdefault}{bx}{n}

\title{
PharmacoNet: Accelerating Large-Scale \\ Virtual Screening by Deep Pharmacophore Modeling
}

\author{%
    Seonghwan Seo \\
    Department of Chemistry, KAIST\\
    \texttt{shwan0106@kaist.ac.kr} \\
    \And
    Woo Youn Kim \\
    Department of Chemistry, KAIST\\
    AI Institute, KAIST\\
    HITS Inc.\\
    \texttt{wooyoun@kaist.ac.kr} \\
}

\begin{document}

\maketitle

\begin{abstract}
As the size of accessible compound libraries expands to over 10 billion, the need for more efficient structure-based virtual screening methods is emerging.
Different pre-screening methods have been developed for rapid screening, but there is still a lack of structure-based methods applicable to various proteins that perform protein-ligand binding conformation prediction and scoring in an extremely short time.
Here, we describe for the first time a deep-learning framework for structure-based pharmacophore modeling to address this challenge.
We frame pharmacophore modeling as an instance segmentation problem to determine each protein hotspot and the location of corresponding pharmacophores, and protein-ligand binding pose prediction as a graph-matching problem.
PharmacoNet is significantly faster than state-of-the-art structure-based approaches, yet reasonably accurate with a simple scoring function.
Furthermore, we show the promising result that PharmacoNet effectively retains hit candidates even under the high pre-screening filtration rates.
Overall, our study uncovers the hitherto untapped potential of a pharmacophore modeling approach in deep learning-based drug discovery.
\end{abstract}

\section{Introduction}
In the realm of modern drug discovery, the search for novel and effective drug molecules has necessitated the exploration of vast chemical spaces.
Structure-based virtual screening with large chemical libraries has attracted great attention as a new paradigm for early drug discovery as make-on-demand libraries of drug-like molecules have recently expanded into the tens of billions \citep{lyu2019ultra, stein2020virtual,gorgulla2020open,luttens2022ultralarge, gentile2022artificial}.
Recent work reported that increasing the chemical library size from 115 million to 11 billion significantly improved the hit rate from 15\% to 30\% \citep{sadybekov2022synthon}.

Molecular docking is a common approach for structure-based virtual screening.
Conventional docking methods involve two steps: generating a large number of binding poses for each initial ligand conformation and ranking binding poses using a well-defined scoring function \citep{friesner2004glide, trott2010autodock, koes2013smina}.
As the scoring functions are based on the physical energy equation, the highest score among the binding poses -- \textit{docking score} -- has been considered as a criterion for virtual screening.
Nevertheless, the pose space is large, continuous, and contains numerous local optima, requiring binding pose search processes spanning from 1 to 100 seconds for each initial conformation \citep{stark2022equibind, lu2022tankbind, corso2022diffdock}.
Considering the virtual screening uses over 100 initial conformers for accurate evaluation, the computational demands of molecular docking become a bottleneck in exploring a wider screening space of billions of molecules \citep{lyu2019ultra, gentile2022artificial}.
This highlights that more efficient structure-based virtual screening approaches are urgently needed.

\newpage

\begin{figure}
  \centering
  \includegraphics[width=1.0\textwidth]{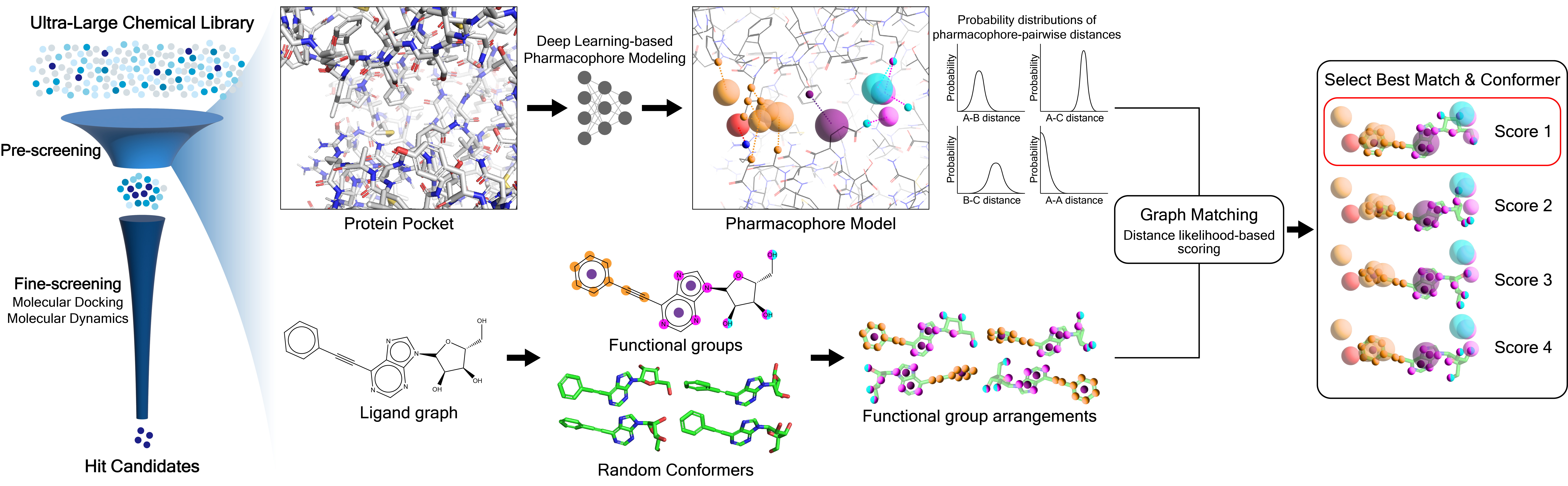}
  \caption{
    Overview of PharmacoNet.
  }
  \label{fig: scheme}
\end{figure}

There are two main approaches for accelerating large-scale virtual screening: the efficient exploration strategy and the pre-screening strategy.
Efficient exploration strategies concentrate on effective molecule selection for molecular docking instead of brute force ligand evaluation \citep{wang2023reducing}.
Recent studies by \citet{sadybekov2022synthon} using the structured library, Enamine REAL \citep{grygorenko2020enamine}, and \citet{graff2021accelerating} employing active learning efficiently explore molecular libraries.
These strategies focus on exploring promising regions within the vast compound space, reducing computational costs associated with library size expansion.

While these exploration strategies reduce the computational burden by bypassing a substantial portion of the library, this also means that they do not perform any evaluation for most of the molecules.
In contrast, the \textbf{pre-screening \& fine-screening} strategy evaluates all molecules \citep{luo2021structure}.
It begins with a preliminary assessment (pre-screening) of the entire library using fast evaluation methods, followed by a more careful assessment (fine-screening) of retained molecules using computationally intensive methods such as molecular docking and molecular dynamics simulations \citep{zhang2023deepbindgcn}.
However, pre-screening methods often rely on preliminary filters with low accuracy, which can lead to the omission of promising \textit{hit candidates} in large-scale virtual screening, where high filtration rates are imperative.
This underscores the importance of the judicious development of pre-screening methods capable of effectively retaining hit candidates at high filtration rates while ensuring rapid processing speed and adaptability to diverse protein targets.

To challenge this issue, we focused on structure-based pharmacophore modeling, a fast structure-based pre-screening approach.
A pharmacophore\footnote{IUPAC \citep{iupac} definition: ``an ensemble of steric and electronic features that is necessary to ensure the optimal supramolecular interactions with a specific biological target and to trigger (or block) its biological response"}
is an abstract description of molecular features with the potential to form the non-covalent interaction (NCI) which is critical for stabilizing protein-ligand complexes, such as aromatic rings and hydrogen bond (H-bond) donors \& acceptors.
Structure-based pharmacophore modeling identifies the optimal 3D pharmacophore arrangement -- \textit{pharmacophore model} -- that a ligand should have to ensure optimal supramolecular interactions \citep{iupac, schaller2020next}.
It can be carried out by determining \textit{protein hotspots}, which are considered as the key protein functional groups to protein-ligand binding \citep{barillari2008hot}, and the optimal location of the pharmacophores to form NCI with each protein hotspot.
Pharmacophore modeling simplifies the protein-ligand interaction prediction by streamlining the complicated topology of the protein-ligand complex structure at the pharmacophore level and representing numerous atom-atom pairwise interactions with a few NCIs.
Ligands are then evaluated based on their structural similarity to the pharmacophore model, enabling rapid and rational screening without atom-level binding pose prediction or energy scoring.

Structure-based pharmacophore modeling divides into the complex-based approach \citep{wolber2005ligandscout, chen2006pocketv2} and the protein-based approach \citep{kirchhoff2001sbp, tran2018allinone, jiang2020autoph4} according to the use of the binding complex conformation of known active ligands.
The protein-based approaches perform pharmacophore modeling solely through structural data of the protein binding site, making them applicable in more general situations, even when the active ligand is unknown.
However, most protein-based approaches heavily rely on expert intuition and occasionally require resource-intensive procedures like molecular dynamics, molecular docking, or fragment crystal structures \citep{yang2010pharmacophore}.
While several automated protein-based approaches \citep{tran2018allinone, jiang2020autoph4} have emerged, they are designed for the fine-screening process, leaving a lack of fully automated protein-based approaches for pre-screening.

\begin{figure}
  \centering
  \includegraphics[width=1.0\textwidth]{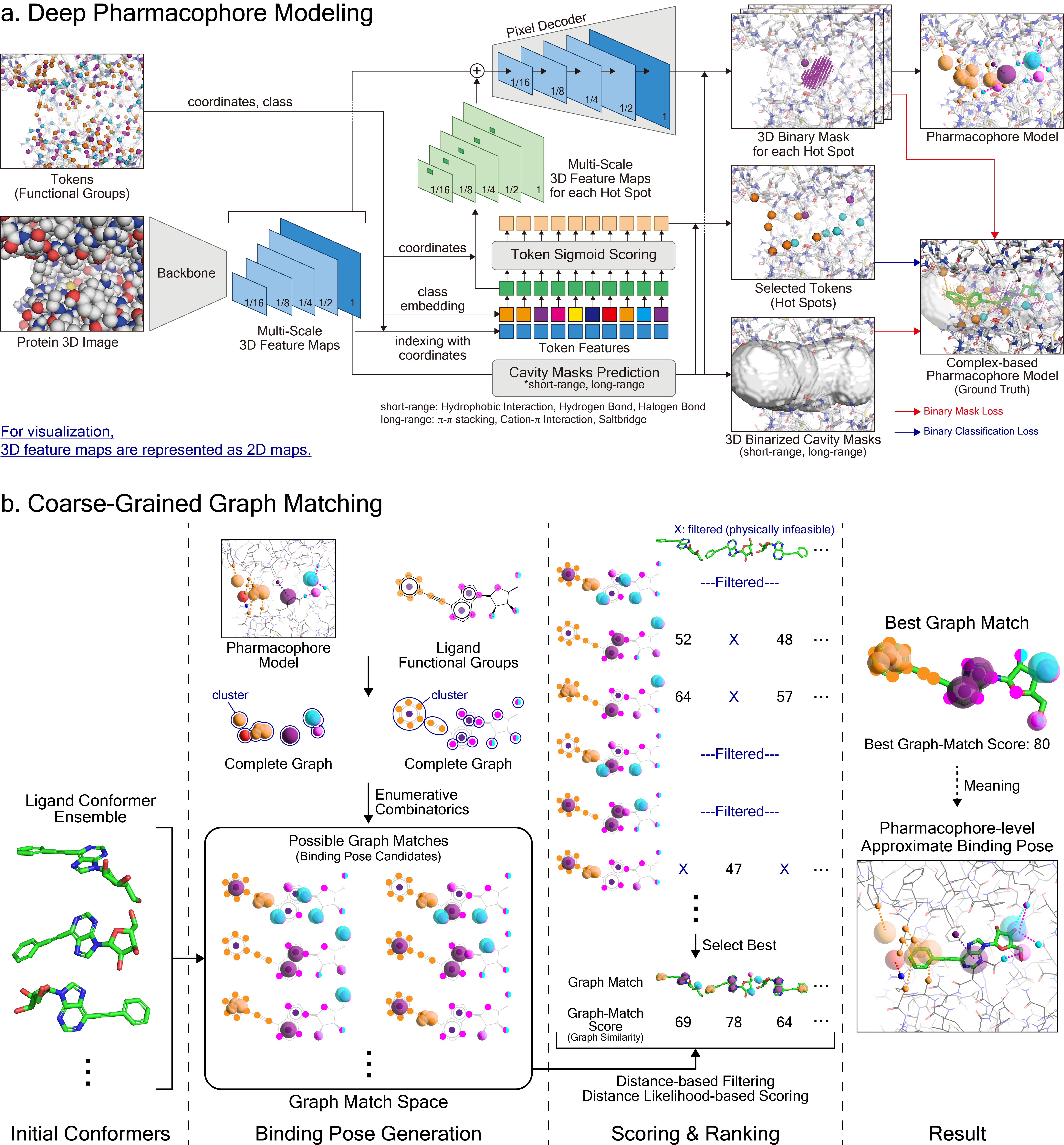}
  \caption{
    The detailed scheme of PharmacoNet.
    \textbf{a.} The architecture of deep learning model for fully automated protein-based pharmacophore modeling.
    For visualization, 3D feature maps are represented as 2D maps.
    The complex-based pharmacophore model is constructed from the crystal structure of the protein-ligand binding complex. (Section \ref{section: pharmacophore_modeling})
    \textbf{b.} The graph-matching algorithm for inexact graph matching.
    The numbers in the figure are arbitrary values. (Section \ref{section: graph_matching})
  }
  \label{fig: model}
\end{figure}

To this end, we proposed PharmacoNet, the first deep-learning framework for structure-based pharmacophore modeling, as illustrated in Figure \ref{fig: scheme} and \ref{fig: model}.
First, it performs fully automated protein-based pharmacophore modeling through framing into image instance segmentation deep learning modeling.
Then, it performs coarse-grained graph matching between the pharmacophore model and the ligand at the pharmacophore level to predict the binding pose.
Finally, we introduce a novel scoring function based on pharmacophore pairwise distance likelihood instead of atom pairwise interactions between the protein and ligand.
PharmacoNet is extremely fast yet reasonably accurate and efficiently retains hit candidates during the pre-screening process.
In this work, PharmacoNet evaluated a million molecules for pre-screening potential inhibitors for KRAS-G12C in 11 minutes on 16 CPU cores of Intel Xeon Gold 6234.

\section{PharmacoNet Architecture}
The high-level protocol of our framework is described in Figure \ref{fig: scheme}.
The model architecture used in PharmacoNet for pharmacophore modeling is shown in Figure \hyperref[fig: model]{\ref*{fig: model}a}, Section \ref{section: pharmacophore_modeling}, and Appendix \ref{appendix: architecture}.
The model training and training dataset is detailed in Appendix \ref{appendix: model_training}.
The graph matching algorithm is described in Figure \hyperref[fig: model]{\ref*{fig: model}b}, Section \ref{section: graph_matching} and Appendix \ref{appendix: graph_matching}.
The scoring function is explained in Section \ref{section: scoring_function}.
The code is available at \url{https://github.com/SeonghwanSeo/PharmacoNet}.

\subsection{Deep learning-based pharmacophore modeling} \label{section: pharmacophore_modeling}
\subsubsection{Instance image segmentation for pharmacophore modeling}
Similar to our approach, \citet{skalic2019ligvoxel} developed LigVoxel, a 3D CNN-based deep learning model for inpainting the chemical functionality map from the pocket image.
However, LigVoxel generates only one map for each of the three functional types (aromatic, H-bond donor, and H-bond acceptor), so it is impossible to recognize individual pharmacophoric features or ensure supramolecular interactions with the binding site.
In the pharmacophore modeling, it is essential to identify the nature of each pharmacophore, including its coordinates, counterpart protein functional groups, and NCI between them \citep{iupac}.
For the identification of individual pharmacophores and their properties, we frame pharmacophore modeling as an image instance segmentation problem, as shown in Figure \hyperref[fig: model]{\ref*{fig: model}a}.

Image segmentation divides an image into segments based on either categories or instance-level criteria \citep{long2015fcn_seg, he2017maskrcnn}.
Semantic segmentation classifies pixels into categories so that the different objects can be grouped into a single segment, similar to LigVoxel.
In contrast, instance segmentation achieves the recognition of both the segment and the category for each object by following procedures: 1) object detection, 2) bounding box delineation, 3) category classification, and 4) binary mask (segment) prediction \citep{he2017maskrcnn}.
In the context of tailoring to pharmacophore modeling, the key differences are:

\begin{enumerate}[topsep=0.5ex,itemsep=0.5ex]
    \item
    Instead of detecting objects from images, our deep learning model selects protein hotspots (instances) among protein functional groups (tokens) in the binding site.
    When a single functional group can form multiple types of NCIs, it is treated as multiple tokens.
    
    \item
    Each instance already contains an NCI type (class).
    Moreover, the possible region (bounding box) for pharmacophore can be obtained from prior knowledge of the maximum length of NCI.
    Therefore, the deep learning model does not predict its class and bounding box.
    
    \item
    A single voxel may belong to multiple instances.
    
    \item
    The deep learning model also addresses an image inpainting problem because the space in the binding site is empty.
    Therefore, there is no definitive answer for the segmentation.
\end{enumerate}

PharmacoNet completes the pharmacophore modeling process in seconds on a single NVIDIA RTX 3090 GPU.
Compared to molecular graph representation, 3D image representation requires more computational cost than molecular graph-based modeling but is advantageous for representing the probability density map of pharmacophores on a grid.
Given that pharmacophore modeling occurs only once for a protein binding site before the virtual screening process, this computational demand is not a significant issue.

\subsubsection{Ground truth: complex-based pharmacophore model} \label{section: groundtruth}
To establish the ground truth for the protein-based pharmacophore modeling, we use the complex-based pharmacophore model obtained from the crystal structures of protein-ligand complexes in the PDBBind v2020 dataset \citep{liu2015pdbbind}.
We considered the protein-ligand functional group pairs that form NCIs as the pairs of the protein hotspot and the pharmacophore.
The identification of NCIs was performed using the Protein-Ligand Interaction Profiler (PLIP) \citep{salentin2015plip}.
For pharmacophore modeling, we used the 7 pharmacophore types: hydrophobic carbon, aromatic ring, H-bond donor \& acceptor, halogen atom, cation, and anion.

Typically, the length of NCIs is less than 6.0~\AA, so it is unnecessary to consider regions far from the pocket cavity when predicting protein hotspots or pharmacophores.
We delineate two cavity regions based on the length of the NCI: one within 5.0~\AA\ from the ligand atoms in crystal structure for short-range interactions (hydrophobic interaction, H-bond, and halogen bond) and another within 7.0~\AA\ for long-range interactions ($\pi$--$\pi$ stacking, cation--$\pi$ interaction, salt bridge).

\subsection{Coarse-grained graph matching} \label{section: graph_matching}
To predict the binding pose of a ligand in a protein, PharmacoNet finds the matching pattern between the ligand and the pharmacophore model.
The ligands can be abstracted at the pharmacophore level by representing the molecular graph of a ligand in terms of functional groups instead of atoms.
Then the pharmacophore model and the ligand functional group arrangement can be represented as 3D complete graphs, so the binding pose prediction problem can be treated as an inexact graph-matching problem.
Since inexact graph matching allows for some tolerance between the correspondence between nodes and edges in two graphs, there are many possible graph matches (PGMs).
To select the most appropriate graph match, a scoring function that reflects the number of matched nodes and the error in the length of edges is required.
Inexact graph matching can be performed between a pharmacophore model and the ligand based on the scoring function and the score of the top-scoring graph match is \textit{graph-matching score} which means the graph similarity.

\paragraph{Ligand functional group arrangements.}
The complete graph of the ligand is denoted as $\mathcal{G}^L = (\mathcal{V}^L, \mathcal{E}^L)$, where each node $v^l_i \in \mathcal{V}^L$ represents the ligand functional group and it contains the set of its possible pharmacophore types $\mathbf{T}^l_i$.
The edge is denoted as $e^l_{ij} \in \mathcal{E}^L$.
For ligand conformer $\mathcal{C}$, the position of node $v^l_i$ is $\mathbf{x}^l_i(\mathcal{C})$, and the length of edge $e^l_{ij}$ is $\zeta(e^l_{ij}, \mathcal{C}) = ||\mathbf{x}^l_i(\mathcal{C}) - \mathbf{x}^l_j(\mathcal{C})||$.

\paragraph{Pharamcophore model.}
The complete graph of the pharmacophore model $\mathcal{G}^P = (\mathcal{V}^P, \mathcal{E}^P)$ contains pharmacophores as nodes, and each node $v^p_i \in \mathcal{G}^P$ has a token score $\mathbf{s}^p_i$, a center position $\mathbf{x}^p_i$, and a radius $\mathbf{r}^p_i$ obtained from Equation \ref{eq: token_score}, \ref{eq: center}, and \ref{eq: radius} in Appendix \ref{appendix: architecture}, with a pharmacophore type $\mathbf{t}^p_i$.
For each edge $e^p_{ij} \in \mathcal{E}^P$, the mean and covariance of length are $\mu(e^p_{ij}) = ||\mathbf{x}^p_i - \mathbf{x}^p_j||$ and $\sigma(e^p_{ij}) = \sqrt{(\mathbf{r}^p_i)^2 + (\mathbf{r}^p_j)^2}$, respectively.

\paragraph{Clustering.}
In the ligand molecules, one functional group often contains numerous functional groups.
For example, 1 aromatic ring such as benzene contains 6 hydrophobic carbons.
To improve the efficiency of the graph matching, we perform clustering for the ligand graph.
The set of clusters is $\mathbb{C}^L$ where each cluster $C^l_i \in \mathbb{C}^L$ is a set of $v^l_{i'}$.
In addition, we perform clustering for the pharmacophore model because one ligand functional group can form interactions with multiple protein functional groups, i.e. one $v^l$ can be matched with multiple $v^p$.
The set of clusters is $\mathbb{C}^P$ where each cluster $C^p_i \in \mathbb{C}^P$ is a set of $v^p_{i'}$.
Graph matching is done on a per-cluster basis, not a per-node basis.
The clustering algorithm is described in Appendix \ref{appendix: clustering}. 

\paragraph{Graph-matching algorithm.}
We can represent a PGM $\mathcal{M}$ as the set of pairs of ligand cluster and pharmacophore model cluster, $\{(C^l_{i_1}, C^p_{j_1}), (C^l_{i_2}, C^p_{j_2}), ...\}$.
In the context of inexact graph matching, the number of all PGMs is $(|\mathbb{C}^P|+1)^{|\mathbb{C}^L|}$, and the computational requirements increase exponentially with the complexity of the ligand or pharmacophore model.
Therefore, the efficient graph-matching algorithm is mandatory.

Indeed, most PGMs are physically infeasible.
For example, it is not reasonable for two ligand functional groups that are only 1~\AA\ to correspond to two pharmacophores that are 10~\AA.
To account for this, we use the distance constraint between cluster pairs as detailed in Appendix \ref{appendix: distance_constraint}.
Furthermore, an ensemble of ligand conformers shares a common core structure.
Across the various conformers, the feasible PGM patterns for the core structure satisfying the distance constraints show substantial similarities.
As a result, graph matching for numerous conformers can be performed simultaneously by identifying a PGM pattern for their core structure and then performing graph matching for the remaining part.
To this end, we implemented the algorithm inspired by the depth-first search algorithm, details of which can be found at \url{https://github.com/SeonghwanSeo/PharmacoNet}.

\subsection{Scoring function} \label{section: scoring_function}
Recently, \citet{shen2023genscore} reported the state-of-the-art scoring model based on the pairwise atom distance likelihood between protein and ligand instead of scoring in the unit of energy.
This has the advantage of allowing relative comparisons between ligands without the need for mapping to experimentally measured affinity values.
Similarly, we propose a distance likelihood-based scoring function to score and rank the PGMs.

A Gaussian Mixture Model is used to express the probability density function of the distance between the ligand graph nodes.
For the two cluster pairs $(C^l_{i_1}, C^p_{j_1}), (C^l_{i_2}, C^p_{j_2}) \in \mathcal{M}$, the probability density function of the distance $d$ between $v^l_a \in C^l_{i_1}$ and $v^l_b \in C^l_{i_2}$ are follows:
\begin{align} \label{eq: mdn}
    & I^{\mathcal{M}}_a = 
        \{s|s \in \{1,2,..., |\mathcal{V}^P|\}, v_s^p \in C^p_s, \mathbf{t}^p_s \in \mathbf{T}^l_a \} \nonumber \\
    & I^{\mathcal{M}}_b = 
        \{t|t \in \{1,2,..., |\mathcal{V}^P|\}, v_t^p \in C^p_t, \mathbf{t}^p_t \in \mathbf{T}^l_b \} \nonumber \\
    & P(d | \mathcal{M}, e^l_{ab}) = F^{\mathcal{M}}_{ab}(d) =
        A
        \sum_{s \in I^{\mathcal{M}}_a}
        \sum_{t \in I^{\mathcal{M}}_b}
        {\pi_{st}}
        \mathcal{N}(d|\mu(e^p_{st}),\sigma(e^p_{st})^2)
\end{align}
where $A$ is the normalizing constant and the weight of each component is the $\pi_{st} = \mathbb{W}[\mathbf{t}^p_s] \mathbb{W}[\mathbf{t}^p_t] \mathbf{s}^p_s \mathbf{s}^p_t$.
The weights for each pharmacophore type $\mathbb{W}[t]$ are in Appendix \ref{appendix: parameters}.
Note that the probability density function $F^{\mathcal{M}}_{ab}$ is independent of the ligand 3D conformer $\mathcal{C}$.

The score of PGM $\mathcal{M}$ for ligand conformer $\mathcal{C}$ can be represented as the weighted sum of pharmacophore pairwise distance likelihood:
\begin{align} \label{eq: score}
    \displaystyle
    &s_{\text{edge}}(\mathcal{C}| \mathcal{M}, e^l_{ab}) =
        \begin{cases}
            w^{\mathcal{M}}_{ab}
            F^{\mathcal{M}}_{ab}(\zeta (e^l_{ab}, \mathcal{C})) 
                & \text{if}~I^{\mathcal{M}}_a \ne \varnothing, I^{\mathcal{M}}_b \ne \varnothing
            \\
            0 
                & \text{else}
        \end{cases}
    \nonumber
    \\
    &\text{Score}(\mathcal{C}, \mathcal{M}) = 
        \sum_{e^l_{ab} \in \mathcal{E}^L} s_\text{edge}(\mathcal{C}| \mathcal{M}, e^l_{ab})
\end{align}
where $w^\mathcal{M}_{ab} = \sum_{s \in I^{\mathcal{M}}_a}{\mathbb{W}[\mathbf{t}^p_s] \mathbf{s}^p_s}/|I^{\mathcal{M}}_a| \times \sum_{t \in I^{\mathcal{M}}_b}{\mathbb{W}[\mathbf{t}^p_t] \mathbf{s}^p_t}/|I^{\mathcal{M}}_b|$.

\section{Experiments}
\subsection{Benchmark Study}
\paragraph{Virtual screening benchmark.} 
The widely-adopted virtual screening benchmarks are CASF-2016 \citep{su2018casf}, DUD-E \citep{mysinger2012dude}, and DEKOIS2.0 \citep{bauer2013dekois}.
However, CASF-2016 only contains 285 molecules and is not well-suited as a benchmark for the pre-screening task.
In contrast, both the DUD-E and DEKOIS2.0 benchmarks are more similar to real-world scenarios.
The DUD-E benchmark comprises 102 targets and offers several hundred actives, along with 50 decoys for each active, all sharing similar physicochemical properties.
Likewise, DEKOIS2.0 contains 81 distinct targets, with 40 actives and a substantial 1200 decoys for each target.
Consequently, our benchmark tests were conducted on both DUD-E and DEKOIS2.0 test sets.

We compared PharmacoNet with state-of-the-art conventional docking methods such as GLIDE (commercial) \citep{friesner2004glide}, AutoDock Vina \citep{trott2010autodock}, and SMINA \citep{koes2013smina}.
In addition, the comparison with the non-structure-based deep learning model for pre-screening, DeepBindGCN \citep{zhang2023deepbindgcn}, is in Appendix \ref{appendix: deepbindgcn}.

\paragraph{Runtime benchmark.}
We followed the benchmark proposed by \citet{stark2022equibind}, which measures the average runtime on 363 complexes discovered since 2019 from PDBBind v2020.
Along with the methods in the conventional docking programs, we also include structure-based deep learning models for binding pose prediction, GNINA \citep{mcnutt2021gnina}, EquiBind \citep{stark2022equibind}, TANKBind \citep{lu2022tankbind}, and DiffDock \citep{corso2022diffdock}. (Appendix \ref{appendix: runtime_benchmark})

\paragraph{Evaluation metric.}
The screening power is assessed with the top $\alpha \%$ enrichment factor, EF$_{\alpha \%}$, which is defined as the ratio of the percentages of the active ligands in the top $\alpha \%$ predictions to the percentages of the active ligands in the total library, and AUROC, the average area under the receiver operating characteristic.
For runtime measurements, the processes that are performed before the virtual screening process are not considered, such as preprocessing of protein or generation of initial conformers for ligand. 

\paragraph{Inference details.}
We generated ligand conformers with RDKit ETKDG \citep{landrum2006rdkit, riniker2015better} from the SMILES and used the top graph matching score among the conformers for screening.
As the default exhaustiveness of Autodock Vina and SMINA is 8, we compared the screening performance of our model when using a single conformer and 8 conformers.

\paragraph{Result.}

\begin{table}
    \caption{
        Results of the benchmark study.
        The top half contains conventional molecular docking methods, and the middle and bottom half contain our framework.
        The number after our model means the number of RDKit conformers used for scoring.
        We reused the number reported in \citet{shen2022rtmscore} and \citet{sunseri2021virtual} for accuracy and \citet{corso2022diffdock} for runtime.
        We ran our model in a 16 CPU-cores environment with Intel Xeon Gold 6234.
        `*' indicates that the method runs on a single CPU core.
        PharmacoNet-U means that the scoring function does not use weights for pharmacophore types.
        PharmacoNet-R means that our deep learning model is trained with a restricted training set.
    }
    \label{tab: vs_benchmark}
    \centering
    \begin{tabular}{lcccc|c|c}
        \toprule
                            & \multicolumn{4}{c|}{DEKOIS2.0}                        & DUD-E         & PDBBind v2020 \citep{stark2022equibind}\\
        Methods             & EF$_{0.5\%}$  & EF$_{1\%}$    & EF$_{5\%}$    & AUROC & EF$_{1\%}$    & 16-CPU Runtime (s)\\
        \midrule            
        GLIDE (comm.)       & 14.6          & 12.5          & 6.30          & 0.747 & 23.6          & 1405* \\
        AutoDock Vina       & 5.46          & 4.51          & 2.82          & 0.633 & 9.93          & 205   \\
        SMINA               & 4.62          & 4.04          & 2.72          & 0.627 & 9.57          & 126   \\
        \midrule
        PharmacoNet (8)     & 3.75          & 3.42          & 2.41          & 0.624 & 5.73          & \textbf{0.0024}\\
        PharmacoNet (1)     & 3.43          & 3.23          & 2.55          & 0.625 & 5.36          & \textbf{0.0018}\\
        \midrule
        PharmacoNet-U (8)   & 1.57          & 1.73          & 2.04          & 0.603 & 4.19          & 0.0024 \\
        PharmacoNet-R (8)   & 3.53          & 3.07          & 2.54          & 0.622 & 4.78          & 0.0024 \\
        \bottomrule
    \end{tabular}
\end{table}

Table \ref{tab: vs_benchmark} highlights that our model is significantly faster than existing energy-based docking methods and deep learning-based models for binding pose prediction.
Our method outperforms the fastest conventional docking program, SMINA, by a factor of 50,000.
To achieve this speed-up, our framework uses a coarse-grained graph-matching algorithm that predicts pharmacophore-level binding poses implicitly without considering atomic-level modeling, as illustrated in \hyperref[fig: model]{\ref*{fig: model}b}.
In addition, our scoring function has only 7 parameters, in contrast to conventional scoring functions that rely on numerous parameters.
Despite these simplifications, our framework exhibits reasonable accuracy through pharmacophore-level abstraction.

\paragraph{Analysis: runtimes by the number of conformers.}
Furthermore, it is worth noting that the computational cost does not increase significantly when using multiple ligand conformers.
In this experiment, we used either 1 or 8 conformers generated by ETKDG for each ligand.
However, real-world virtual screening scenarios \citep{lyu2019ultra, sadybekov2022synthon} often involve multi-conformer databases containing tens to hundreds of conformers for each ligand, typically generated by fast conformer generation software such as OMEGA \citep{hawkins2010omega}.
Since our graph-matching approach is similar to exhaustive rigid-body docking methods \citep{vigers2004rigid4, mcgaughey2007rigid3}, which uses numerous ligand conformers to account for the flexibility of the ligand structure, so using more conformers can improve accuracy.
To assess the efficiency of our framework in handling large numbers of conformers, we measured the runtime according to the number of ETKDG conformers, as shown in Appendix \ref{appendix: runtime}.
The average runtime for 100 conformers is 8.9 ms, while for a single conformer is 1.8 ms, and this result indicates the effectiveness of our framework in real-world scenarios.

\paragraph{Analysis: weights for pharmacophore types.} \label{section: parameters}
Our scoring function relies solely on the weights assigned to each pharmacophore type as its parameters.
In this study, these weights were determined based on prior knowledge of the relative contribution of each pharmacophore to protein-ligand binding as described in Appendix \ref{appendix: parameters}.
To investigate the factors affecting the performance of our framework, we performed an ablation study (PharmacoNet-U) using the DEKOIS2.0 benchmark and the DUD-E benchmark used in previous experiments.
Table \ref{tab: vs_benchmark} shows that the accuracy of our scoring function is improved by reflecting the chemistry of protein-ligand binding.
Given that the current scoring function only uses 7 hand-designed parameters for pharmacophore types, higher performance can be reached by using additional parameters such as atom type with parameter fitting.

\paragraph{Analysis: generalization ability of deep learning model.}
In this work, we trained our model with PDBBind v2020 \citep{liu2015pdbbind} dataset which contains 19,443 molecules.
Since the limited number of structural data relative to the topological complexity of the protein-ligand complexes, A common challenge of deep learning approaches in drug discovery is the low generalization ability to proteins not encountered during training \citep{moon2022pignet}.
To address this, we simplify the topology of the protein-ligand complex through coarse-grained modeling at the pharmacophore level and incorporate distance constraints based on the maximum length information of NCIs.
To improve generalization ability, we incorporated pharmacology as an inductive bias into the deep learning model and tailored the image instance segmentation architecture to pharmacophore modeling.

In the evaluation of the generalization capabilities of our model, we conducted experiments (PharmacoNet-R) using a restricted dataset consisting of 5000 training complexes and 500 validation complexes randomly selected from the original training and validation sets, representing 30\% of the PDBBind v2020 dataset.
Table \ref{tab: vs_benchmark} shows that the model trained on this restricted training set has a slightly lower overall performance.
Nevertheless, it remains competitive in terms of AUROC and performs better than the baseline in EF$_{5\%}$.
This result demonstrates that the model can learn overall pharmacological features even with a limited training set, indicating that our approach achieves high generalization ability.

\subsection{Pre-screening for Large-Scale Virtual Screening}
To assess the effectiveness of our model's pre-screening capabilities, we performed a large-scale virtual screening of 1 million ChEMBL molecules for KRAS (G12C), a challenging drug target.
One of the key objectives in the pre-screening is to retain the hit candidates, the molecules with the best docking scores.
Note that only the top 100 to 1000 molecules in the entire library are considered hit candidates for experimental validation.
In evaluating the pre-screening performance of our framework, we used the top-$N$ enrichment factor, denoted EF$_{N}$, which is defined as the ratio of the percentages of hit candidates in the retained molecules to the percentages of these candidates in the entire library, when the hit candidates are considered as the top $N$ molecules with the highest docking score in the entire library.
Furthermore, we measured the average docking score in conjunction with the enrichment factor.
For the pre-screening stage, we used the top graph matching score among the 10 RDKit ETKDG conformers for each ligand, and for the fine-screening stage, we used SMINA with default settings.

\begin{table}
    \caption{
        Results of the pre-screening.
        `Filtration rate' and `Number of retained' indicate the filtration rates of pre-screening and the number of remaining molecules after pre-screening, respectively.
        `Total time' is the time spent on the entire virtual screening including pre-screening (PharmacoNet) and fine-screening (SMINA).
        The last line represents the baseline scenario without any pre-screening.
    }
    \label{tab: prescreening}
    \centering
    \begin{tabular}{cc|ccc|c|c}
        \toprule
        Filtration      & Number of     & \multicolumn{3}{|c|}{KRAS}                    & Average            & Total time (hours)\\
        rate            & retained      & EF$_{100}$    & EF$_{500}$    & EF$_{1000}$   & docking score      & 16-CPU\\
        \midrule                                 
        90\%            & 100,000       & 5.46          & 4.86          & 4.76          & -8.35              & 150 \\
        95\%            & 50,000        & 8.53          & 7.74          & 7.50          & -8.43              & 75 \\
        98\%            & 20,000        & 9.92          & 9.33          & 9.58          & -8.50              & 31 \\
        99\%            & 10,000        & 15.9          & 12.5          & 12.0          & -8.55              & 15 \\
        \midrule
        -               & 992,235       & -         & -         & -                 & -7.66              & 1300 \\
        \bottomrule
    \end{tabular}
\end{table}

\begin{figure}
     \centering
     \begin{subfigure}[b]{0.30\textwidth}
         \centering
         \includegraphics[width=\textwidth]{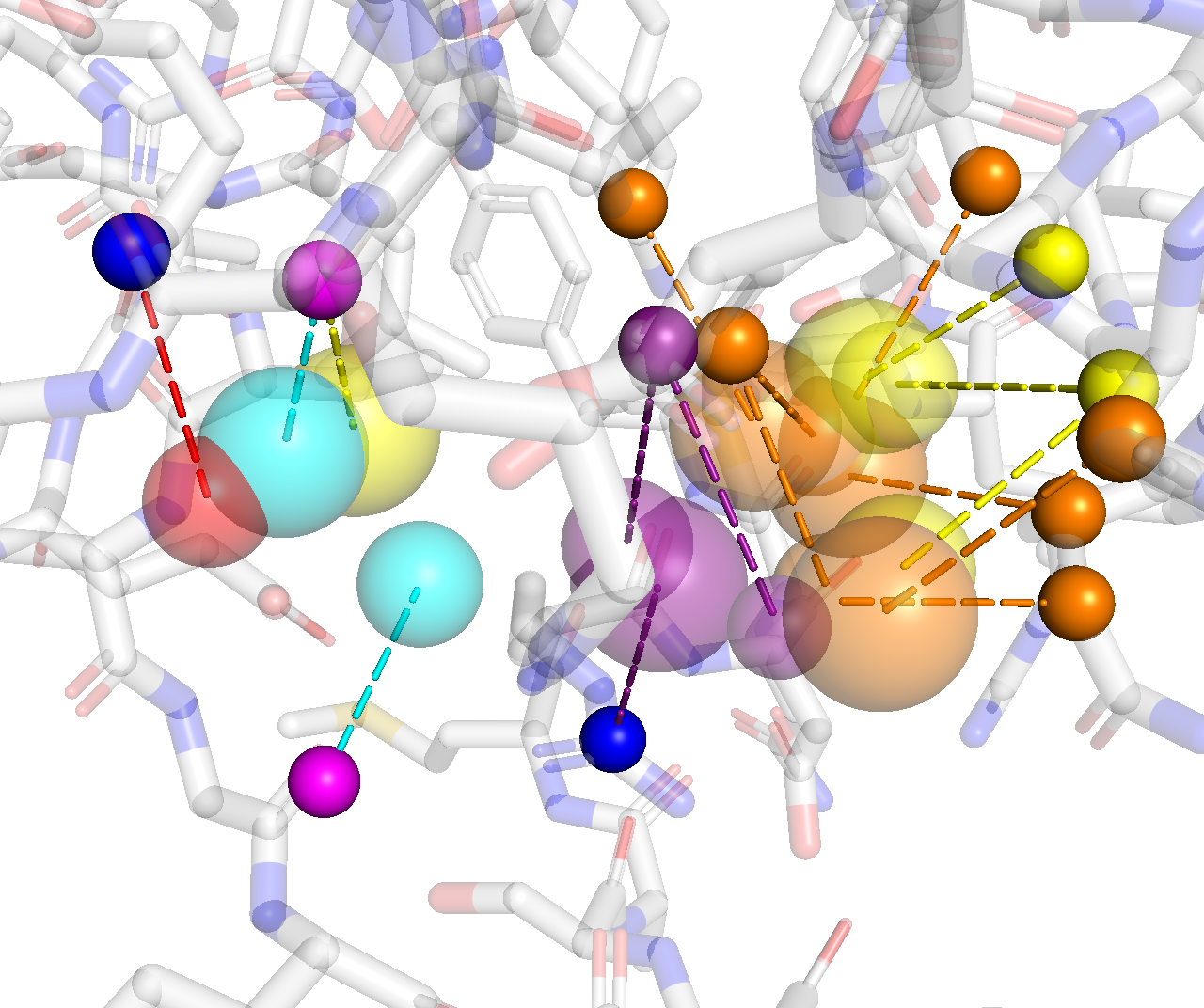}
         \caption{Pharmacophore model}
     \end{subfigure}
     \hfill
     \begin{subfigure}[b]{0.38\textwidth}
         \centering
         \includegraphics[width=\textwidth]{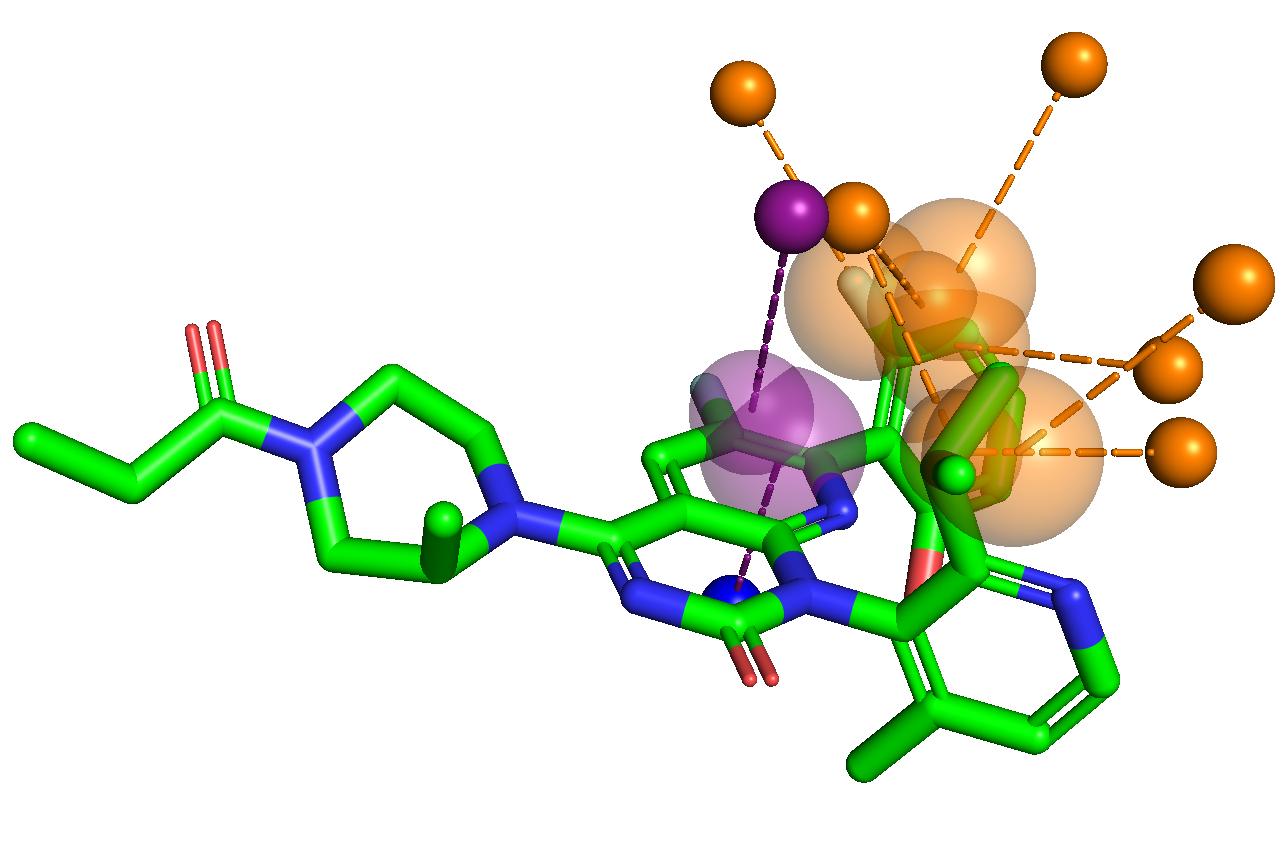}
         \caption{AMG-510, crystal structure}
     \end{subfigure}
     \begin{subfigure}[b]{0.28\textwidth}
         \centering
         \includegraphics[width=\textwidth]{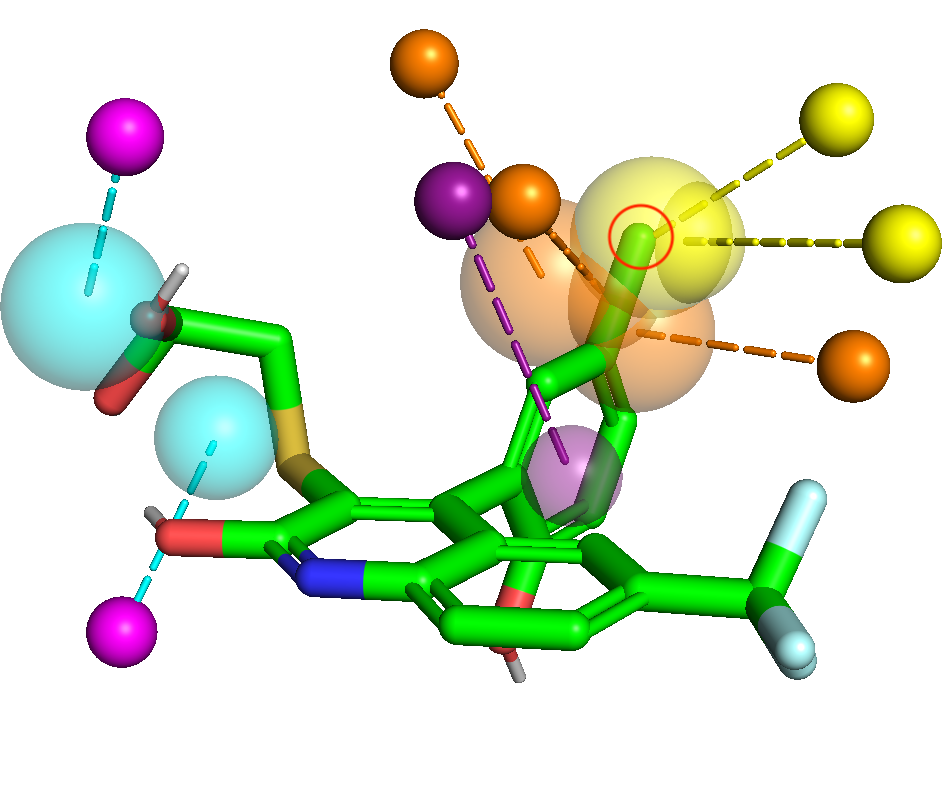}
         \caption{CHEMBL183526}
     \end{subfigure}
     \hfill
        \caption{
            Visualizations of the pharmacophore model for KRAS.
            The color of pharmacophore and hot spot is as follows: orange for hydrophobic carbon, purple for aromatic ring, red for anion, blue for cation, cyan for H-bond donor, magenta for H-bond acceptor, and yellow for halogen atom and halogen bond acceptor.
            (a) The generated pharmacophore model of the binding site (PDB ID: 6OIM).
            (b) The crystal structure of the known inhibitor (AMG-510). 
            (c) The SMINA docking pose of the ligand with the highest score of pre-screening.
        }
        \label{fig: visualize}
\end{figure}

The pre-screening process was remarkably efficient, taking only \textbf{11 minutes for 1 million molecules} on 16 CPU cores.
The results in Table \ref{tab: prescreening} show that our model can significantly reduce the total virtual screening time while effectively retaining hit candidate molecules.
By filtering out 90\% of the molecules from the library, our model retained more than half of the top 100 hit candidates (0.01\% of the library), and 16 of those with a 99\% filtration rate.
Moreover, it's worth noting that increasing the filtration rate of pre-screening improved the average docking scores of the remaining molecules.
These observations suggest that there is a trend between our scores and energy-based docking scores, even though energy is not considered in model training.

Figure \ref{fig: visualize} shows the generated pharmacophore model for KRAS.
Interestingly, in the SMINA docking pose of CHEMBL183526, which achieved the highest matching score, the chlorine atom (indicated by a red circle) aligns with the halogen atom pharmacophores (represented by yellow spheres) in our model, demonstrating the ability of our model to identify pharmacophore sites.
Notably, the corresponding halogen bond was absent from both the X-ray (PDBID: 6OIM) and cryo-EM (PDBID: 8G47) structures of AMG-510, a known inhibitor of KRAS.
This indicates the potential of our model to reveal unknown pharmacophores contributing to energy stabilization, which are not present in the complex structure of the active ligand.

\subsection{Limitations}
PharmacoNet relies on a scoring and graph-matching algorithm at the pharmacophore level, which has reasonably accurate and strong generalization capabilities without the need for a detailed parameter-fitting procedure.
However, this approach has a trade-off; it does not capture atomic-level features.
For example, it cannot distinguish between variations in the strength of the same salt bridge due to differences in charge and atom type, nor does it account for intramolecular energy.

To address these limitations, it may be useful to integrate force-field-based approaches such as UFF \citep{rappe1992uff} and MMFF \citep{halgren1996mmff} and atomic-level details into the graph-matching and scoring algorithm, where a parameter-fitting process becomes essential.
In particular, a promising avenue for improvement is the integration of a machine learning or deep learning-based approach into the graph-matching algorithm.
This integration will significantly enhance our framework's ability to account for fine-grained details and account for intramolecular energy factors, ultimately leading to improved accuracy and performance.

Furthermore, designing a balanced scoring model for both binding pose prediction and screening has long been a challenge in the molecular docking task \citep{shen2023genscore, moon2023pignet2}.
The success of the re-scoring models \citep{shen2023genscore, moon2023pignet2, moon2022pignet, shen2022rtmscore} based on docking pose indicates that our framework has the potential to improve accuracy by integrating additional scoring functions tailored to each task, rather than relying on a single function for both.

\section{Conclusion}
In this work, we have demonstrated for the first time the ability and potential of a deep learning approach based on pharmacophore modeling.
Our novel model, PharmacoNet, a 3D image instance segmentation model for generating pharmacophore models, opens up fresh avenues of the structure-based deep learning approach by coarse-grained modeling of proteins and ligands at the pharmacophore level.
With a coarse-grained graph-matching algorithm and a simple distance likelihood-based scoring function, our framework achieves significant speed-up with reasonable accuracy and high generalization ability.
Moreover, our framework has revealed its potential as a pre-screening method in the realm of large-scale virtual screening.
We believe that the incorporation of pharmacophore modeling and pharmacophoric knowledge into deep learning-based methods paves the way for innovative breakthroughs in drug discovery.

\begin{ack}
This work was supported by Basic Science Research Programs through the National Research Foundation of Korea (NRF), grant funded by the Ministry of Science and ICT (NRF-2023R1A2C2004376, RS-2023-00257479).
\end{ack}
\newpage

\section{Supplementary Material}
\subsection{Detailed Model Architecture} \label{appendix: architecture}
\subsubsection{Atomic features}
For each atom in the binding site, we use the residue type (20 common amino acids and unknown for others), atom type (C, N, O, S, and unknown for others), and functional group type (hydrophobic carbon, H-bond donor \& acceptor, halogen bond acceptor, aromatic ring, cation, and anion) as an atomic feature.
All water molecules and metal ions are omitted.

\subsubsection{Voxelization of binding site}
We represent the protein binding site as a 3D voxelized image.
In this work, we used an image size of $64\times64\times64$ with a resolution of $0.5$~\AA\ to represent the binding site, resulting in a box side length of $32$~\AA.
Note that the side length of the recommended maximum search box of AutoDock Vina is $30$~\AA.

For the protein binding site $\mathcal{B}$, the coordinates and atomic features of the binding site atom are denoted $\mathbf{x}^b_i \in \mathbb{R}^3$ and $\mathbf{h}^b_i \in \mathbb{R}^C$, respectively.
Then, the 3D input image $\mathsf{I} \in \mathbb{R}^{D \times H \times W \times C}$ is represented as follows:
\begin{align}
    \mathsf{I}_{d, h, w, :} &= \sum_i^{N^b}{K(|| \mathcal{T}^{\mathsf{I} \rightarrow \mathcal{B}} ([d, h, w]) - \mathbf{x}^b_i ||) \times \mathbf{h}^b_i} \\
    K(x) &=
    \begin{cases}
        e^{-2x^2}   &~\text{if}~x \le 1.5 (\text{\AA})\\
        0           &~\text{else}
    \end{cases}
\end{align}
where $K$ is a kernel function, $N^b$ is the number of atoms in binding site, $D$, $H$, $W$ are the spatial dimensions, and $C$ is the number of atomic features.
$\mathcal{T}^{A\rightarrow B}$ is the mapping between the voxel coordinates system and the real-world coordinates system, e.g. $\mathcal{T}^{\mathsf{I} \rightarrow \mathcal{B}} ([d, h, w])$ is the real-world coordinates of a voxel with coordinates of $[d, h, w]$.

\subsubsection{Protein hotspot prediction}
To detect protein hotspots, PharmacoNet uses the Feature Pyramid Network \citep{lin2017feature} with a 3D extension of the Swin Transformer V2 Encoder \citep{liu2022swin}, which is typically used in object detection tasks.
The network obtains multi-scale 3D feature maps:
\begin{equation}
    \mathbb{F} = \{\mathsf{F}^{(1)}, \mathsf{F}^{(1/2)}, ...\} = \phi^{\text{backbone}}(\mathsf{I})
\end{equation}
where $\mathsf{F}^{(s)} \in \mathbb{R}^{sD \times sH \times sW \times C_s}$ is the feature map for each scale $s$ and corresponding hidden dimension $C_s$.

Each protein functional group (token), denoted as $t_i^t$, contains its coordinates $\mathbf{x}_i^t$, and its NCI type (class) $c_i^t$.
Since some functional groups can form multiple types of interactions, our deep learning model distinguishes them with their interaction type.
The score of each token for determining whether it is a hotspot or not is then calculated as follows:
\begin{align}
    &d_i, h_i, w_i = \mathcal{T}^{\mathcal{B} \rightarrow \mathsf{F}^{(1)}}(\mathbf{x}^t_i)  \\
    &\mathbf{z}^t_i = \phi^{\text{token}}(\mathsf{F}^{(1)}_{d_i,h_i,w_i,:}, c^t_i) \\
    &y^t_{i} = \text{sigmoid}(\phi^{\text{score}}(\mathbf{z}^t_i))
    \label{eq: sigmoid_score} \\
    &\mathbf{s}^t_{i} = P(X^{\text{score}}_{c^t_i} < y^t_i)
    \label{eq: token_score}
\end{align}
where the $X^{\text{score}}_{c^t_i}$ is the random variable of the sigmoid score ($y^t_i$) distribution of all tokens within the cavity in the validation set. 
As the number of protein hotspots constitutes only a small fraction of all tokens, we use the token score's relative position within the score distribution, $\mathbf{s}^t_{i}$, instead of directly using the sigmoid score in the scoring function for PharmacoNet.
The score distribution and threshold for each NCI type are in Appendix \ref{appendix: hotspot}.

\subsubsection{Binary mask prediction for each cavity}
\label{section: prediction_cavity}
We defined two types of cavities based on the length of the NCI.
Each cavity is obtained as follows:
\begin{align}
    \mathsf{C}^\text{long}  &= \text{sigmoid}(\phi^{\text{cavity}}_\text{long}(\mathsf{F}^{(1)}))
    \label{eq: longcavity}\\
    \mathsf{C}^\text{short} &= \text{sigmoid}(\phi^{\text{cavity}}_\text{short}(\mathsf{F}^{(1)}))
    \label{eq: shortcavity}
\end{align}
where $\mathsf{C}^\text{long}, \mathsf{C}^\text{short} \in \{0, 1\}^{D \times H \times W}$ represent binarized outputs with a threshold of 0.5 for the cavity regions corresponding to long-range and short-range interactions, respectively.

\subsubsection{Binary mask prediction for each protein hotspot}
\label{section: prediction_segmap}
In the instance segmentation manner, our model predicts the optimal positions of the pharmacophore for each protein hotspot:
\begin{equation}
    \mathsf{D}  = \text{sigmoid}(\phi^{\text{mask}}(\mathbb{F}, \mathbf{x}^t_i, \mathbf{z}^t_i)) \odot \mathsf{C}^{\text{short}}
\label{eq: densitymap}
\end{equation}
where the $\mathsf{D} \in [0, 1]^{D \times H \times W}$ is the density map, and the binary mask is obtained from $\mathsf{D}$ with the threshold of 0.5.
We perform the Gaussian smoothing with the kernel size of 5 and the sigma of 0.5 to reduce the noise in the model output.
Then, the voxels within 1~\AA\ from protein atoms are masked, since NCIs are much longer than 1.0~\AA.
Moreover, the radius of the bounding box according to pharmacophore type is as follows:
\begin{itemize}
    \item 4.5~\AA\ for hydrophobic interaction, halogen bond, and H-bond
    \item 6.0~\AA\ for $\pi$--$\pi$ stacking and salt bridge
    \item 6.5~\AA\ for cation-$\pi$ interaction
\end{itemize}
Each radius is longer than the maximum distance in the default hyperparameters of Protein-Ligand Interaction Profiler (PLIP) \citep{salentin2015plip}.

\subsubsection{Pharmacophore modeling from segmentation map}
The pharmacophores for each protein hotspot can be represented as the segments in a corresponding binary mask.
When there are multiple segments in a single binary mask, each segment is considered a distinct pharmacophore.
Since the segment is the group of voxels, the center $\in \mathbb{R}^3$ and radius $\in \mathbb{R}$ of the corresponding pharmacophore are defined as follows:
\begin{align}
    \text{center}  &= 
        \frac{1}{\sum_n^N {\mathsf{D}_{d_n, h_n, w_n}}}
        \sum_n^N {\mathsf{D}_{d_n, h_n, w_n} \mathcal{T}^{\mathsf{D} \rightarrow \mathcal{B} }\left( [ d_n, h_n, w_n ]\right)}
    \label{eq: center}
    \\
    \text{radius}  &= \sqrt[3]{N/(4\pi/3)} \times \text{resolution}
    \label{eq: radius}
\end{align}
where $N$ is the number of containing voxels, $d_n, h_n, w_n$ is the coordinates of each voxel in the segment, and $\mathsf{D}^i$ are the model output and binary mask corresponding to the pharmacophore.

\subsection{Deep learning model training and inference} \label{appendix: model_training}
\paragraph{Dataset.}
We used the PDBbind v2020 dataset \citep{liu2015pdbbind}, a collection of high-resolution crystal structures, and measured binding affinities for 19,443 protein-ligand complexes deposited in the Protein Data Bank (PDB) \citep{berman2000pdb}.
Following \citet{shen2023genscore}, we partitioned the dataset into 17,658 training complexes and 1,500 validation complexes, excluding 285 CASF-2016 \citep{su2018casf} test complexes.
We omitted 8 complexes from the training set that contained multiple ligands.
The center of mass of the active ligand was taken as the center of the binding site, and random translations and rotations were applied for data augmentation.

\paragraph{Objective function.}
From the complex-based pharmacophore model which is the ground truth pharmacophore model, we set the sigmoid scores for the protein hotspots to 1, otherwise 0.
Each pharmacophore is voxelized into a 3D binary mask for its hotspot, where voxels within 1.0~\AA\ are set to 1, otherwise 0.
The cavities are also voxelized to 3D binary masks.
The dimensions of the ground truth 3D binary masks are $(D, H, W)$.

For the instance segmentation modeling, we utilize two loss terms, which are pixel-wise binary mask loss and binary classification loss.
Both loss terms employ the binary cross entropy loss function between ground truth and deep learning model prediction obtained from Equation \ref{eq: sigmoid_score}, \ref{eq: longcavity}, \ref{eq: shortcavity} and \ref{eq: densitymap}. (Appendix \ref{appendix: architecture})

\paragraph{Training details}
The prediction of binary masks for each pharmacophore is carried out using a 3D-CNN which is computationally expensive.
As a result, we train the model by sampling up to 6 pharmacophores for each complex in each training iteration.
Furthermore, the number of protein hotspots (instances) is considerably smaller compared to the total number of functional groups (tokens) present within the binding site.
Consequently, we sample up to 200 tokens for each complex, with priority given to protein hotspots, tokens within cavities, and remaining tokens to reduce the data imbalance.

For the Swin Transformer V2-based encoder, we used the SwinV2-T configuration with the exception of the patch size of 2 and the window size of 4.
The total number of parameters of our deep learning model is 31M.
We use AdamW optimizer \citep{loshchilov2017adamw} for 200,000 iterations with a batch size of 8, the learning rate of $1\times10^{-4}$, weight decay of 0.05, and $(\beta_1, \beta_2)$ of $(0.9, 0.999)$.
The model training takes about 48 hours on 4 NVIDIA RTX 3090 GPUs.

\paragraph{Inference details.}
The voxel outside the maximum distance of the NCI (based on the PLIP rule) from the corresponding hot spot is masked with 0.
In addition, the voxels within 1.0~\AA~ of all protein atoms are masked.
Finally, the model performed gaussian smoothing with a kernel size of 5 and a sigma of 0.5 to reduce the noise in the segmentation map.

\subsection{Graph-Matching Algorithm.} \label{appendix: graph_matching}
All notations are defined in Section \ref{section: graph_matching}.

\subsubsection{Parameters for scoring function} \label{appendix: parameters}
In this study, these weights were determined based on prior knowledge of the relative contribution of each pharmacophore to protein-ligand binding.
Hydrogen and halogen bonds were considered comparable in terms of their characteristics and strengths, with both significantly outweighing the influence of hydrophobic interactions.
Furthermore, salt bridges were considered stronger than both hydrogen and halogen bonds, while $\pi$--$\pi$ stacking was recognized as a predominant driving force contributing to complex stability, particularly in terms of entropy.
Based on this prior knowledge, we set the weight as follows:
\begin{equation*}
    \mathbb{W}[t] = \begin{cases}
        1   & \text{if } t \text{ is Hydrophobic carbon}  \\
        4   & \text{if } t \text{ in } \{\text{H-bond donor, H-bond acceptor, Halogen atom}\}  \\
        8   & \text{if } t \text{ in } \{\text{Cation, Anion, Aromatic ring}\}
    \end{cases}
\end{equation*}

\subsubsection{Clustering algorithm} \label{appendix: clustering}
\paragraph{Ligand functional groups.}
We perform clustering for ligand functional groups according to the following criteria:
\begin{itemize}
    \item The cations, anions, H-bond acceptors, and H-bond donors in the same functional group are grouped.
    \item The aromatic ring and hydrophobic carbons in the same functional group are grouped.
    \item The hydrophobic carbons in the same carbon chain are grouped.
\end{itemize}

\newpage
\paragraph{Pharmacophore model.}
Algorithm \ref{algorithm: model_clustering} describes the clustering algorithm for the pharmacophore model.
\begin{algorithm}[!hb]
	\caption{Pharmacophore Model Graph Clustering Algorithm}
    \label{algorithm: model_clustering}
	\begin{algorithmic}[1]
        \State UsedNodes $\leftarrow \{\}$ 
        \State $\mathbb{C}^P \leftarrow \{\}$ 
		\For {$v^p_i$ in $\mathcal{G}^P$}
            \If {$v^p_i$ in UsedNodes}
                \State \textbf{continue}
            \EndIf
            \State $C^p \leftarrow \{v^p_i\}$ 
            \If {$t^p_i$ in \{Cation, Anion, Aromatic\} }
        		\For {$v^p_j \ne v^p_i$ in $\mathcal{G}^P$}
                    \If {$t^p_i = t^p_j$ and $||\mathbf{x}^p_i-\mathbf{x}^p_j|| < 1.5$\AA}
                		\State add $\{v^p_j\}$ to $C^p$
                    \ElsIf {$||\mathbf{x}^p_i-\mathbf{x}^p_j|| < 3.0$\AA}
                        \If {$t^p_i$ in \{Cation, Anion\} and $t^p_j$ in \{H-bondDonor, H-bondAcceptor\}}
                    		\State add $v^p_j$ to $C^p$
                        \ElsIf {$t^p_i$ is Aromatic and $t^p_j$ is HydrophobicCarbon}
                    		\State add $v^p_j$ to $C^p$
                        \EndIf
                    \EndIf
                \EndFor
            \EndIf
            \State add $C^p$ to $\mathbb{C}^P$
            \State UsedNodes $\leftarrow$ UsedNodes $\cup~C^p$
        \EndFor
		\For {$v^p_i$ in $G^P$}
            \If {$v^p_i$ in UsedNodes}
                \State \textbf{continue}
            \EndIf
            \State $C^p \leftarrow \{v^p_i\}$ 
            \If {$t^p_i$ in \{HydrophobicCarbon, HalogenAtom\} }
        		\For {$v^p_j \ne v^p_j$ in $\mathcal{G}^P$}
                    \If {$t^p_i = t^p_j$  and $||\mathbf{x}^p_i-\mathbf{x}^p_j|| < 3.0$\AA}
                		\State add $\{v^p_j\}$ to $C^p$
                    \EndIf
                \EndFor
            \ElsIf {$t^p_i$ in \{H-bondDonor, H-bondAcceptor\} }
        		\For {$v^p_j \ne v^p_j$ in $\mathcal{G}^P$}
                    \If {$t^p_j$ in \{H-bondDonor, H-bondAcceptor\} and $||\mathbf{x}^p_i-\mathbf{x}^p_j|| < 3.0$\AA}
                		\State add $\{v^p_j\}$ to $C^p$
                    \EndIf
                \EndFor
            \EndIf
            \State add $C^p$ to $\mathbb{C}^P$
            \State UsedNodes $\leftarrow$ UsedNodes $\cup~C^p$
        \EndFor
	\end{algorithmic} 
\end{algorithm} 

\newpage

\subsubsection{Distance-based filtering algorithm} \label{appendix: distance_constraint}
\begin{algorithm}[!hb]
	\caption{Distance-based Filtering Algorithm}
    \label{algorithm: distance_constraint}
	\begin{algorithmic}[1]
        \State \textbf{Input:} $(C^l_{i_1}, C^p_{j_1}), (C^l_{i_2}, C^p_{j_2}), \mathcal{C}$
        \State $N_{\text{total}} = 0$
        \State $N_{\text{pass1}} = 0$
		\For {$v^l_{a_1}$ in $C^l_{i_1}$}
    		\For {$v^l_{a_2}$ in $C^l_{i_2}$}
                \State $N_{\text{pair}} = 0$
                \State $N_{\text{pass2}} = 0$
                \For {$v^p_{b_1}$ in $C^p_{j_1}$}
                    \For {$v^p_{b_2}$ in $C^p_{j_2}$}
                        \If {$t^p_{b_1} \not\in T^l_{a_1}$ or $t^p_{b_2} \not\in T^l_{a_2}$}
                            \State \textbf{continue}
                        \EndIf
                        \State $N_{\text{pair}} \leftarrow N_{\text{pair}} + 1$
                        \If {$|\zeta(e^l_{a_1 a_2}, \mathcal{C}) - \mu(e^p_{b_1 b_2})| \le 2\sigma(e^p_{b_1 b_2})$}
                            \State $N_{\text{pass2}} \leftarrow N_{\text{pass2}} + 1$
                        \EndIf
                    \EndFor
                \EndFor
                \If {$N_{\text{pair}} > 0$}
                    \State $N_{\text{total}} \leftarrow N_{\text{total}} + 1$
                    \If {$N_{\text{pass2}} \ge 0.5 \times N_{\text{pair}}$}
                        \State $N_{\text{pass1}} \leftarrow N_{\text{pass1}} + 1$
                    \EndIf
                \EndIf
            \EndFor
        \EndFor
        \If {$N_{\text{total}} > 0$ and $N_{\text{pass1}} \ge 0.5 \times N_{\text{total}}$}
            \State \textbf{return True}
        \Else
            \State \textbf{return False}
        \EndIf
	\end{algorithmic} 
\end{algorithm} 

\subsection{Software details} \label{appendix: train_inference_detail}
\paragraph{Conformer generation.}
We used the ETKDG \citep{riniker2015better} version 3 (small ring) implemented in RDKit \citep{landrum2006rdkit}.

\paragraph{Protein-ligand interaction profiler.}
To detect non-covalent bonds from the structure of a protein-ligand complex in PDBBind v2020, we used the Protein-Ligand Interaction Profiler (PLIP) \citep{salentin2015plip} and Open Babel \citep{o2011openbabel}.
For flexible and fast data processing during the model training, we reimplemented it while keeping the original PLIP rules except for the water bridge and the metal bridge.

\paragraph{Molecular voxelization.}
In this study, our framework voxelizes the binding sites or pharmacophores using Gaussian and binary kernels.
To accomplish this, we introduce a new voxelization tool, MolVoxel, designed to enable on-the-fly voxelization in various environments.
MolVoxel is implemented in Python and has minimal dependencies, rendering it highly versatile for a wide range of deep learning and machine learning applications.
Currently, it supports NumPy, Numba, and PyTorch (with CUDA support).
The code is available in \url{https://github.com/SeonghwanSeo/molvoxel}.

\newpage

\subsection{Additional Result} \label{appendix: additional}
\subsubsection{Hot Spot Detection} \label{appendix: hotspot}
\begin{figure}[!htb]
  \centering
  \includegraphics[width=1.0\textwidth]{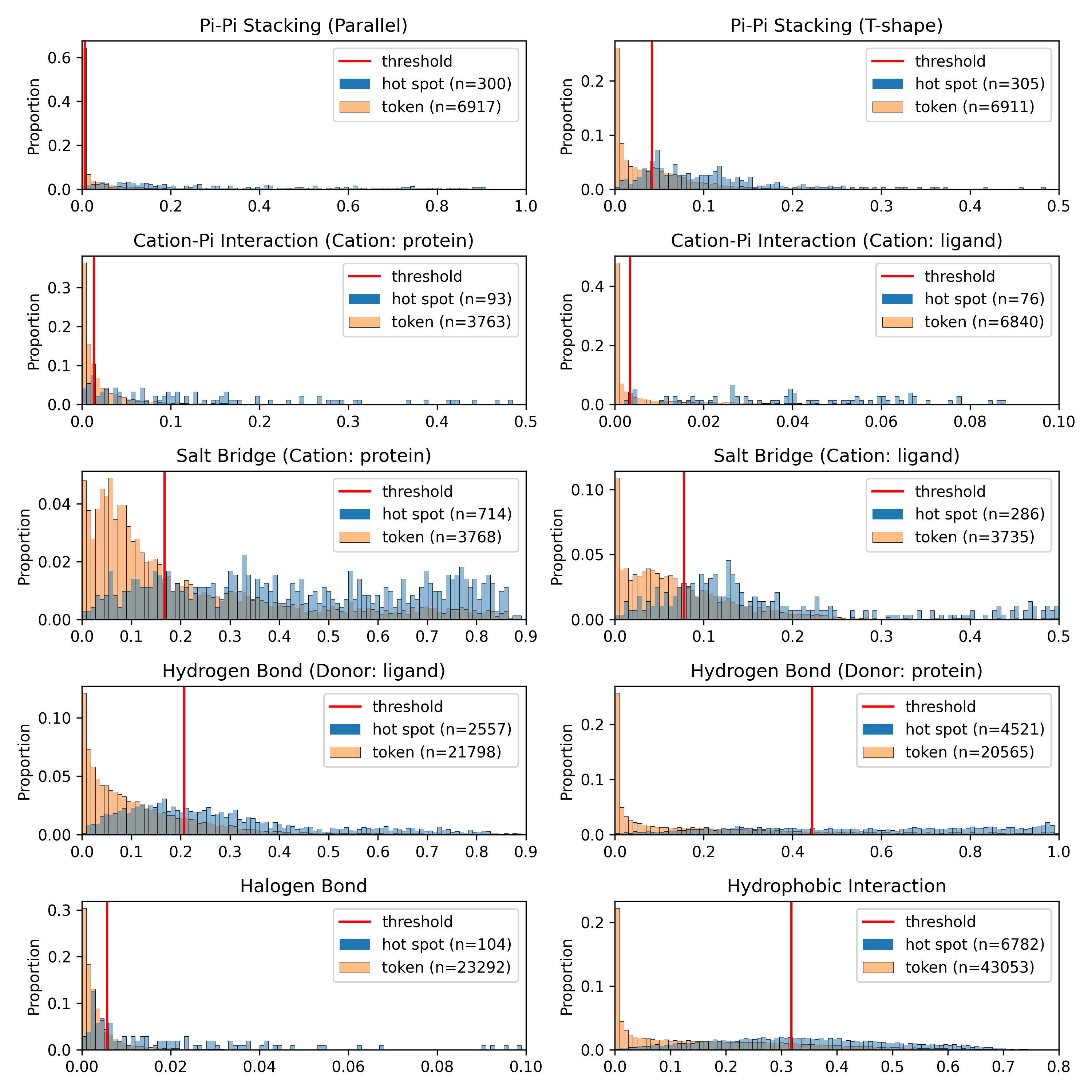}
  \caption{
    The sigmoid score distribution of tokens in the validation set.
    The blue histograms mean the score distribution of hot spots and the orange ones mean the score distribution of all tokens within cavities.
    The numbers in the parentheses mean the number of tokens.
    The red line means the threshold for hot spot prediction.
  } 
  \label{fig: hotspot}
\end{figure}

Figure \ref{fig: hotspot} shows the score distributions and thresholds for non-covalent interaction types.
To determine hot spots, we considered the top 40\% of scores for $\pi$--$\pi$ stacking, cation--$\pi$ interactions, and salt bridges, where the number of tokens is relatively small, and the top 20\% of scores for hydrophobic interactions, H-bonds, and halogen bonds, where the number of tokens is larger.

\newpage
\subsubsection{Comparison with non-structure-based deep-learning model for pre-screening} \label{appendix: deepbindgcn}
\citet{zhang2023deepbindgcn} developed a deep learning model DeepBindGCN for pre-screening that works for multiple targets by integrating the 1D molecular vector of a pocket and a ligand.
The model uses only the information of protein pocket and ligand 2D graph without the information of complex, so the prediction can be performed by combining the pre-calculated vector of ligands and proteins.

However, spatial and topological information is lost during global pooling to represent them as a 1D vector.
Since the spatial coordination of protein and ligand is crucial in determining their interactions, the loss of spatial information leads to low performance and generalization ability for unseen target \citep{moon2022pignet}.
We compare the performance of our model with the DeepBindGCN-RG, the regression version of DeepBindGCN.
Table \ref{tab: compare_deepbindgcn} shows that DeepBindGCN-RG is faster than our model by GPU acceleration, but our model is faster in the CPU environment and shows better accuracy.

\begin{table}[h]
    \small
    \caption{
        Comparison with non-structure-based deep learning model for pre-screening, DeepBindGCN-RG.
        EF$_{0.5\%}$, EF$_{1\%}$, and EF$_{5\%}$ are average top 0.5\%, 1.0\%, 5.0\% enrichment factor, respectively.
        PDBBind is the runtime benchmark where '16-CPU' and 'GPU' mean the average runtime for each device.
        The numbers in parenthesis after our model mean the number of RDKit conformers used.
    }
    \label{tab: compare_deepbindgcn}
    \centering
    \begin{tabular}{lccc|ccc|cc}
        \toprule
                            & \multicolumn{3}{c|}{DEKOIS2.0}                 & \multicolumn{3}{c|}{DUD-E}                           & \multicolumn{2}{c}{PDBBind}\\
        Methods             & EF$_{0.5\%}$  & EF$_{1\%}$    & EF$_{5\%}$     & EF$_{0.5\%}$     & EF$_{1\%}$       & EF$_{5\%}$     & 16-CPU (s)        & GPU (s)  \\
        \midrule                                                               
        DeepBindGCN-RG      & 0.89          & 0.80          & 0.82           & 2.31             & 2.11             & 1.62           & 0.003             & \textbf{0.0005}\\
        PharmacoNet (8)     & \textbf{3.75} & \textbf{3.42} & 2.41           & \textbf{7.23}    & \textbf{5.73}    & 3.23           & 0.0024            & -\\
        PharmacoNet (1)     & 3.43          & 3.23          & \textbf{2.55}  & 6.30             & 5.36             & \textbf{3.41}  & \textbf{0.0018}   & -\\
        \bottomrule
    \end{tabular}
\end{table}

\subsubsection{Comparison with structure-based deep learning models for binding pose prediction} \label{appendix: runtime_benchmark}
Recently, several deep learning models have been developed to predict the protein-ligand complex conformation more accurately or quickly.
To evaluate the efficiency of our coarse-grained graph-matching algorithm, we performed a comparison with the corresponding models.
Table \ref{tab: runtime_benchmark} shows that our method outperforms the fastest conventional docking programs by a factor of 10,000, and it is still 10 times faster than the fastest deep learning model designed for binding pose prediction.

\begin{table}[h]
  \caption{
  Average runtime per prediction of conventional docking programs and several deep learning-based approaches.
  The top five lines contain energy-based binding pose optimization approaches and the middle three lines contain data-driven approaches based on complex crystal structures.
  The last two lines show our models and the number after our model means the number of RDKit conformers used.
  `*' indicates that the method runs on a single CPU core.
  We reused the number from \citet{stark2022equibind}, \citet{lu2022tankbind}, and \citet{corso2022diffdock} for baseline approaches.
  }
  \label{tab: runtime_benchmark}
  \centering
  \begin{tabular}{llcc}
    \hline
                    &                       & \multicolumn{2}{c}{Average Runtime (s)} \\
    Methods         & Type                  & 16-CPU    & GPU  \\
    \hline
    GLIDE (comm.)   & Energy-based binding pose prediction \& scoring & 1405*     & -\\
    AutoDock Vina   &                       & 205       & -\\
    QVina-W         &                       & 49        & -\\
    SMINA           &                       & 146       & -\\
    GNINA           &                       & 247       & 146   \\
    \hline
    EquiBind        & Data-driven binding pose prediction  & 0.16      & 0.04  \\
    TANKBind        &                       & 0.54      & 0.28  \\
    DiffDock        &                       & -         & 10/40 \\
    \hline
    PharmacoNet (8) & Coarse-grained graph matching \& scoring & \textbf{0.0024} & -   \\
    PharmacoNet (1) &                       & \textbf{0.0018} & -   \\
    \hline
  \end{tabular}
\end{table}

\newpage
\subsubsection{Runtime according to the number of Ligand Conformers} \label{appendix: runtime}
We measured runtime using up to 100 conformers to understand how runtime increases with the number of conformers used in the evaluation.
The evaluation is performed on the same runtime benchmark set proposed by \citet{stark2022equibind}.
We attempted to generate ETKDG conformers up to 4,000 times for each ligand.
We excluded 7 RDKit-unreadable molecules and 11 molecules for which fewer than 100 conformers were generated from the result.
The reported time is the average of 5 runs.
All measurements were made in a 16-CPU environment with the same machine, Intel Xeon Gold 6234.
\begin{figure}[!h]
  \centering
  \includegraphics[width=0.8\textwidth]{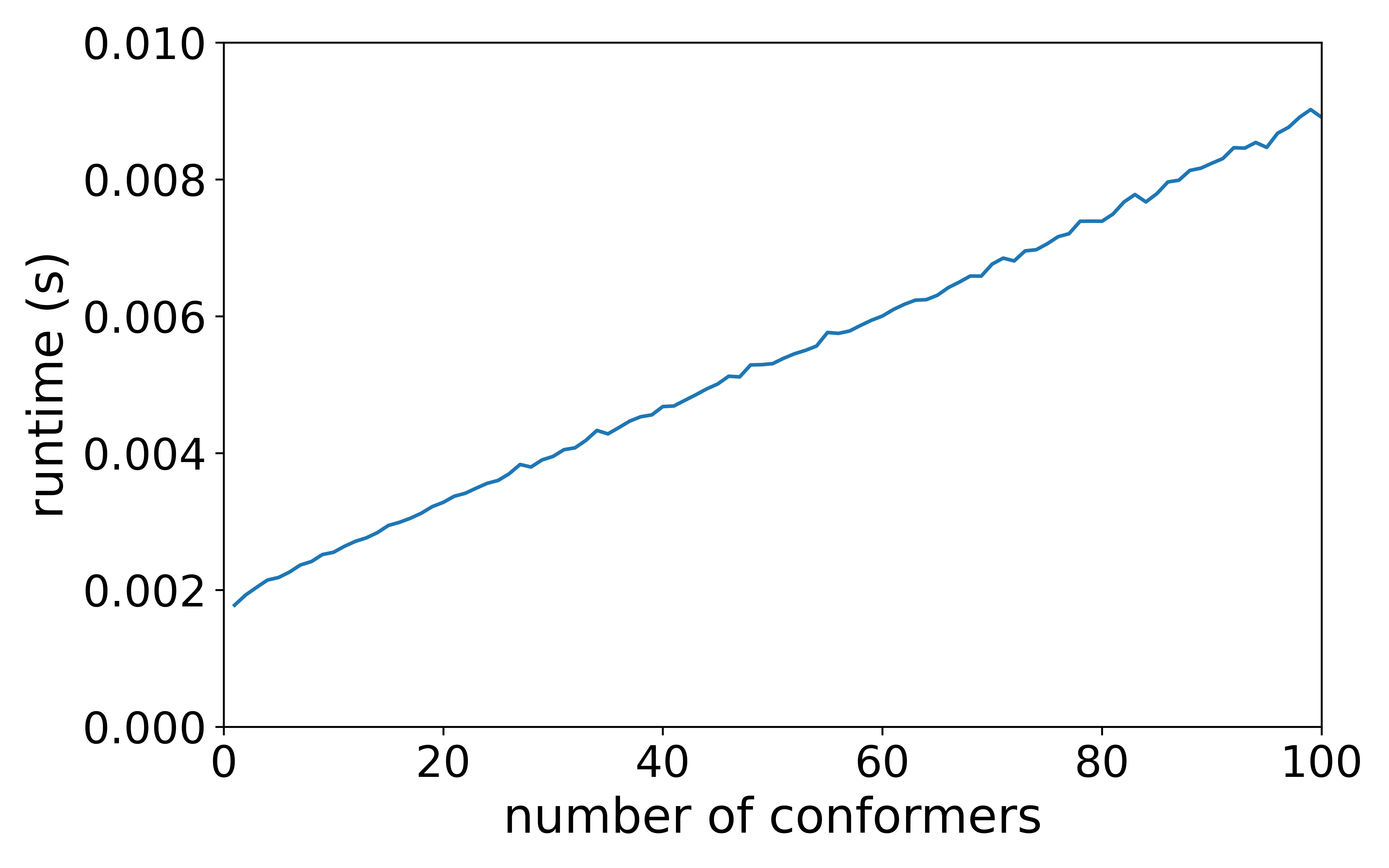}
  \caption{
    Average runtime according to the number of RDKit ETKDG conformations.
  }
  \label{fig: runtime}
\end{figure}

\subsection{Related Work}
\paragraph{Protein-ligand binding pose prediction.}
Traditional binding pose prediction method is based on physical energy-based scoring function and optimization algorithm \citep{friesner2004glide, trott2010autodock, koes2013smina}.
Recent works improve the performance by incorporating deep learning-based scoring function \citep{mcnutt2021gnina, mendez2021deepdock}. 
Another promising approach is to frame the pose prediction task as a regression problem \citep{stark2022equibind, lu2022tankbind} or a generative modeling problem \citep{corso2022diffdock} with deep learning.
EquiBind \citep{stark2022equibind} and TANKBind \citep{lu2022tankbind} directly predict the atomic positions of the binding pose, so they are much faster than the traditional docking methods.
However, their performance has not reached that of traditional docking, and they often cause steric crashes, leading to incorrect scores with physics-based and deep learning-based scoring functions.\citep{corso2022diffdock, yu2023deep}
DiffDock \citep{corso2022diffdock} succeeds in generating high-quality conformations through diffusion generative modeling, but it is not fast enough to be used for pre-screening due to multiple denoising steps.

\paragraph{Rigid exhaustive docking.}
The flexibility of the molecules is one of the biggest challenges to predicting conformations.
It makes optimization difficult by not only increasing the pose space but also creating numerous local optima.
To avoid this problem, one of the main docking strategies is rigid-body docking with an ensemble of ligand conformers \citep{friesner2004glide, vigers2004rigid4, mcgaughey2007rigid3}.
Rigid exhaustive docking is a practical docking strategy with sets of numerous core conformations of ligands \citep{stark2022equibind}.

\paragraph{Machine learning-based scoring function for virtual screening.}
Recently, machine learning-based and deep learning-based scoring functions have been developed to boost virtual screening.
The scoring strategies are usually divided into non-structure-based \citep{zhang2023deepbindgcn, gentile2020deepdocking} and structure-based \citep{shen2023genscore, shen2022rtmscore, moon2022pignet, moon2023pignet2, lim2019predicting} whether they take into account the conformations of protein-ligand complexes.
Non-structure-based scoring methods have the advantage of not requiring the use of complex structures
However, since the spatial relationship between protein and ligand is essential in determining their interaction, structure-based scoring models show better performance and generalization ability by recognizing key interactions from the complex conformation \citep{panday2019silico}.
Therefore, non-structure-based models are usually used in ligand-based virtual screening with known active ligands of the target protein \citep{bahi2018ligVS, sharma2022ligVS_btk}, and structure-based models are used in virtual screening strategies for general targets.

\paragraph{Model generalization to unseen protein.}
The generalization ability to unseen proteins is one of the biggest challenges for structure-based scoring models.
Several studies have pointed out that deep learning models are often trained on the biases inherent in the data instead of the physical interaction between proteins and ligands \citep{shen2023genscore, chen2019hidden}.
To improve the generalization ability, incorporating pharmacological properties as an inductive bias is a promising means in the various tasks of deep learning-based drug discovery \citep{moon2022pignet, moon2023pignet2, zhung2023protein, imrie2021deep}.

\paragraph{Instance image segmentation.}
Image segmentation is the task of partitioning an image into multiple segments.
These segments may be delineated based on either semantic attributes (e.g. person, dog, building) \citep{long2015fcn_seg} or instance-level criteria \citep{he2017maskrcnn}.
Semantic segmentation was tackled as a pixel classification problem, where each pixel is classified into one of the categories.
Therefore, different objects can be grouped into one segment in semantic segmentation.
In contrast, instance segmentation recognizes the segment and category for each object together.
Instance segmentation is divided into a two-stage approach that detects objects and then determines the mask of each object \citep{he2017maskrcnn}, and a one-stage approach that performs segmentation on the entire image and then assigns a category to each segment \citep{cheng2022mask2former, jain2023oneformer}.

\medskip

{
\small
\bibliography{neurips_ai4d3_2023}
}


\end{document}